\documentclass[reprint, superscriptaddress, secnumarabic, amssymb, nobibnotes, aps, prl]{revtex4-1}

\usepackage{chemformula,siunitx}
\usepackage{lastpage}
\usepackage[protrusion=true, expansion=true]{microtype}
\usepackage{wrapfig}
\usepackage{multirow}
\usepackage{array}
\usepackage{ragged2e}
\usepackage{titlesec}
\justifying

\usepackage{graphicx}
\usepackage{epstopdf}
\usepackage[T1]{fontenc}
\usepackage{amsbsy}
\usepackage[T1]{fontenc}
\usepackage{amsmath}
\usepackage{amssymb}
\usepackage{bbm}
\usepackage{braket}
\usepackage{xcolor}
\allowdisplaybreaks
\usepackage{graphicx}
\usepackage[normalem]{ulem}
\DeclareUnicodeCharacter{2212}{-}

\usepackage[colorlinks=true]{hyperref}
\hypersetup{
    unicode=false,          
    pdftoolbar=true,        
    pdfmenubar=true,        
    pdffitwindow=false,     
    pdfstartview={FitH},    
    pdftitle={ZrTe2.8},    
    pdfauthor={Poulami Manna},      
    pdfsubject={},   
    pdfcreator={},   
    pdfproducer={}, 
    pdfkeywords={} {} {}, 
    pdfnewwindow=true,      
    colorlinks=true,       
    linkcolor=blue, 
    citecolor=blue,        
    filecolor=magenta,      
    urlcolor=blue          
}

\begin{document}
\title{Topological and Planar Hall Effect in Monoclinic van der Waals Ferromagnet NbFeTe$_2$}
\author{Suchanda Mondal}
\affiliation{Department of Physics, Indian Institute of Science Education and Research Bhopal,
Bhopal Bypass Road, Bhauri, Madhya Pradesh 462-066, India}
\author{Shubhankar Roy}
\affiliation{Vidyasagar Metropolitan College, 39 Sankar Ghosh Lane, Kolkata 700006, India}
\author{Poulami Manna}
\affiliation{Department of Physics, Indian Institute of Science Education and Research Bhopal,
Bhopal Bypass Road, Bhauri, Madhya Pradesh 462-066, India}
\author{Ravi Prakash Singh}
\email{rpsingh@iiserb.ac.in}
\affiliation{Department of Physics, Indian Institute of Science Education and Research Bhopal,
Bhopal Bypass Road, Bhauri, Madhya Pradesh 462-066, India}
\date{\today}

\begin{abstract}
Two-dimensional (2D) van der Waals (vdW) ferromagnets have emerged as a critical class of quantum materials for next-generation, low-dimensional spintronic devices. In this study, we report a comprehensive study of the transport properties of the layered soft ferromagnet $\text{NbFeTe}_2$. We report the first observation of the topological Hall effect (THE) and the planar Hall effect (PHE) in metallic $\text{NbFeTe}_2$. THE signatures persist up to 45 K, while PHE remains evident well above Curie temperature ($T_C$). The observed negative longitudinal magnetoresistance, along with the PHE, provides strong evidence for a nontrivial electronic band structure. The coexistence of perpendicular magnetic anisotropy and a substantial THE : two key properties that are highly desirable for future spintronics applications make monoclinic vdW ferromagnetic $\text{NbFeTe}_2$ a promising platform to advance spintronics applications.
\end{abstract}

\maketitle
\section{INTRODUCTION}
The isolation of layered van der Waals (vdW) ferromagnets down to the monolayer and bilayer limits, together with the observation of long-range magnetic order at finite temperature, has stimulated intense exploration of emergent phenomena in two-dimensional (2D) magnets \cite{huang2017,klein2019,chen2019,tan2018,gong2017,chen2018,dzhong2017,syan2024,song2018,jiang2018}. In these systems, magnetic anisotropy plays a crucial role by opening a gap in the magnon excitation spectrum, thereby suppressing thermal fluctuations and stabilizing ferromagnetism in reduced dimensions. These advances have reshaped both theoretical understanding and experimental approaches to low-dimensional magnetism. In recent years, particular attention has been directed toward 2D vdW magnets hosting topologically nontrivial spin textures, as their thickness-dependent tunability of spin structure is integral to the development of ultralow-power electronic and spintronic technologies \cite{xlv2024,kkim2018,mbrich2022}. In such systems, the interplay among uniaxial magnetic anisotropy, exchange interactions, and magnetic dipolar interactions stabilizes nontrivial spin textures.

Within this broader context, the layered vdW material NbFeTe$_2$ has emerged as an intriguing platform due to its novel and tunable physical properties \cite{Neuhausen,li1992,wu2024,bai2019}. NbFeTe$_2$ crystallizes in an orthorhombic structure and has been identified as an Anderson insulator exhibiting spin-glass-like behavior with competing intralayer ferromagnetic (FM) and interlayer antiferromagnetic (AFM) interactions \cite{wu2024,bai2019}. Moreover, the intrinsic hopping mechanism of the charge-excitation carriers in NbFeTe$_2$ can be utilized as a promising route for efficient cryogenic THz detection \cite{wang2025}. Recent advances in synthesis have enabled the stabilization of an FM phase of NbFeTe$_2$ with reduced structural symmetry, achieved by careful optimization of the growth temperature. In particular, higher synthesis temperatures favor the formation of a monoclinic phase that exhibits a clear FM metallic behavior \cite{wu2024,stepanova2024}. Monte Carlo simulations suggest that in monoclinic NbFeTe$_2$, nearest-neighbor Fe spins are canted in Nb-deficient regions, pointing to an inherently complex spin structure \cite{gao2025}.

Recently, layered magnetic systems, such as Fe$_n$GeTe$_2$ ($n = 3, 4, 5$), Fe$_3$GaTe$_2$, Cr$_{1+x}$Te$_2$ have been reported to host intricate spin textures, including skyrmion bubble, (anti)meron chain, noncoplanar spin structure, and magnetic skyrmion phase, below their respective transition temperatures \cite{bding2020,YGao2020,czhang2023,yhuang2025}. In addition to the intrinsic, momentum-space Berry phase-driven anomalous Hall effect (AHE), the above systems exhibit the topological Hall effect (THE), which is considered a potential hallmark for the identification of nontrivial spin structures in magnetic systems \cite{spurwar2023,mhuang2021,ywang2019,You2019,Bera2025,Algaidi2025}. Moreover, the presence of large anisotropic magnetoresistance (AMR) and the planar Hall effect (PHE) in Cr$_{1+x}$Te$_2$ flakes induced by spin-dependent scattering underscores their relevance for future sensing and memory devices \cite{xma2023}. Motivated by these developments, we have conducted a comprehensive study of the magnetic and magnetotransport properties of a monoclinic FM single crystal of NbFeTe$_2$.

In this work, we have shown that NbFeTe$_2$ exhibits FM order below $T_C$ near 80 K with an out-of-plane easy axis. Temperature-dependent resistivity reveals contributions from multiple inelastic scattering mechanisms, while magnetoresistance (MR) measurements at different temperatures show a negative longitudinal MR (LMR). Magnetotransport studies further uncover, for the first time, the emergence of THE at low temperatures and a robust PHE that persists well above $T_C$ in metallic NbFeTe$_2$. These results establish NbFeTe$_2$ as an FM metal that hosts a rich spin–charge interplay, with the potential to stabilize nontrivial topological spin textures. Future investigations on atomically thin layers of monoclinic NbFeTe$_2$ may open avenues to engineering its magnetic and topological responses in 2D systems.
 
\section{EXPERIMENTAL DETAILS}
NbFeTe$_2$ single crystals were synthesized using the chemical vapor transport method, taking iodine as the transport agent. A stoichiometric mixture of high-purity Nb, Fe, and Te was sealed in an evacuated quartz tube and placed in a two-zone gradient furnace. The tube was then heated at 800$^{\circ}$C/700$^{\circ}$C with a temperature gradient of 4$^{\circ}$C/cm for 10 days. The single crystals obtained were in the form of shiny silver-colored plates. Before characterization and measurements, the crystals were freshly cleaved to expose clean surfaces. Structural analysis was performed at room temperature by X-ray diffraction (XRD) using a PANalytical diffractometer equipped with monochromatic Cu-$K$$_{\alpha}$ radiation. High-resolution transmission electron microscopy (HRTEM) of the crystals was performed with a FEI TALOS F200S microscope. Magnetization measurements were performed under ambient pressure using a superconducting quantum interference device vibrating sample magnetometer (MPMS 3, Quantum Design) in the temperature range of 2–300 K. Electrical and magnetotransport properties were investigated using a physical property measurement system (PPMS) employing a standard four-probe configuration. Fig. 1(a) shows the crystal structure of NbFeTe$_2$. The as-grown NbFeTe$_2$ crystallizes in a monoclinic phase with the space group $P2_1/c$ (No. 14). The crystal structure comprises layers in which Nb atoms form a distorted honeycomb network, with Fe–Fe dumbbells positioned centrally within each Nb hexagon and sandwiched between the layers of Te atoms [upper panel of Fig. 1(a)]. The XRD pattern obtained from the as-grown facet of the single crystal NbFeTe$_2$, shown in Fig. 1(b), displays sharp and well-defined Bragg reflections, indicative of high crystalline quality and confirming that the exposed surface of the crystals is oriented parallel to the $bc$-plane. In the inset of Fig. 1(b), a typical single crystal of NbFeTe$_2$ of dimension 2 mm$\times$1 mm$\times$0.30 mm is shown. The HRTEM image of a typical NbFeTe$_2$ single crystal is shown in the lower left panel of Fig. 1(a). The lower right panel of Fig. 1(a) illustrates the selected area electron diffraction (SAED) pattern created by the crystallographic planes, confirming the high crystalline quality and the single-crystalline nature of the grown samples. The elemental composition was analyzed using a scanning electron microscope equipped with energy-dispersive X-ray (EDX) spectroscopy. The measured atomic ratios of Fe:Nb:Te were found to be approximately 1.00:0.98:2.02, closely matching the nominal stoichiometry (as shown in Fig. S1(a) in the Supplementary Material). 

 \begin{figure*}
    \centering
    \includegraphics[width=1\linewidth]{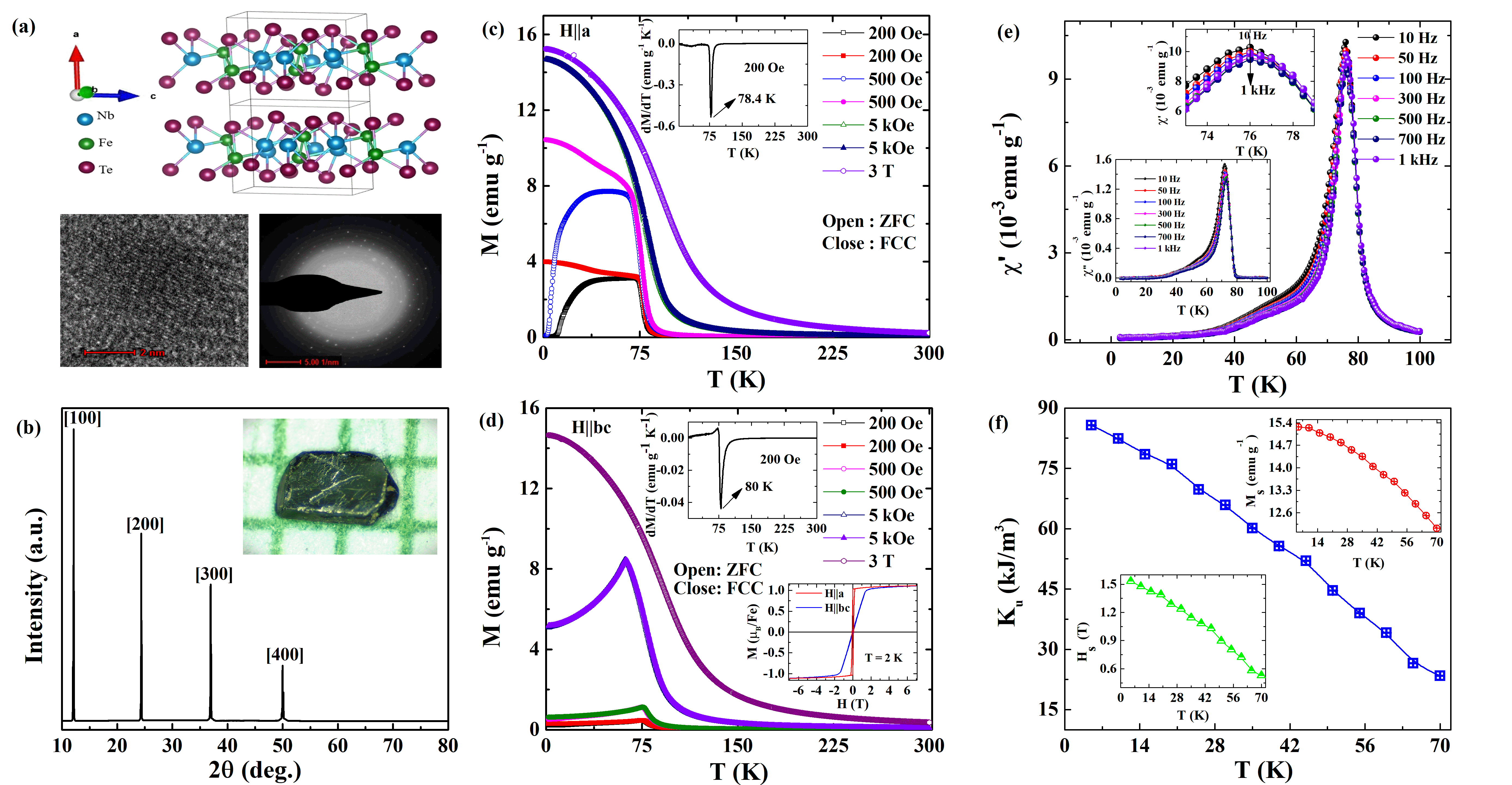}
    \caption{(a) (Upper panel) Crystal structure of monoclinic NbFeTe$_2$, where the sky, green, and violet spheres define the Nb, Fe, and Te atoms, respectively. (Left of the lower panel) The high-resolution transmission electron microscopy (HRTEM) image of the as-grown crystal. (Right of the lower panel) SAED pattern obtained from a single grain with HRTEM measurement. (b) Typical X-ray diffraction pattern obtained from the cleaved plane of NbFeTe$_2$ single crystal at room temperature exhibits exclusively (h00) Bragg reflections. The optical image of the as-grown single crystal is shown in the inset. Temperature dependence of the dc magnetization measured in zero-field-cooled and field-cooled conditions for applied magnetic field in the range 200 Oe to 3 T for (c) $H$ $||$ $a$ and (d) $H$ $||$ $bc$. The inset of (c) and the upper inset of (d) show the first derivative of the $M$-$T$ curve under FC condition for $H$ $||$ $a$ and $H$ $||$ $bc$ directions, respectively. Lower inset of (d) shows isothermal magnetization for both $H$ $||$ $a$ axis and $H$ $||$ $bc$ plane at $T =$ 2 K. (e) The in-phase component of AC magnetic susceptibility $\chi'(T)$ measured at different frequencies ranging from 10 Hz to 1 kHz using 2.5 Oe ac field. The zoomed view of frequency dependence is shown in the upper inset. The temperature dependence of the out-of-phase component of AC magnetic susceptibility $\chi''(T)$ with varying frequencies is shown in the lower inset. (f) Temperature-dependence of the anisotropy constant $K_u$. The temperature variations of $M_s$ and $H_s$ are shown in the upper and lower insets, respectively.}
    \label{fig:enter-label}
 \end{figure*}

\section{RESULTS AND DISCUSSIONS}
 The temperature-dependent magnetization of NbFeTe$_2$ single crystals measured under zero-field-cooled (ZFC) and field-cooled (FC) conditions in fields ranging from 0.02 to 3 T applied along the in-plane ($H$ $||$ $bc$) and out-of-plane ($H$ $||$ $a$) directions is presented in Fig. 1. The curves in Fig. 1(c) and (d) show a notable anisotropic magnetic response below 80 K. A clear enhancement in $M$($T$, $H$) is observed near $T_C$, consistent with the reported paramagnetic (PM) to FM phase transition \cite{wu2024,stepanova2024}. For $H$ $||$ $a$, the ZFC and FC magnetization curves exhibit significant thermomagnetic irreversibility in lower magnetic fields below the transition temperature. Additionally, the ZFC magnetization exhibits a sharp drop near 40 K as the temperature decreases. In contrast, for $H$ $||$ $bc$, both the ZFC and FC curves decrease with temperature below the transition temperature. A similar magnetization downturn is also observed in other layered vdW magnetic systems, i.e., Cr$_2$Y$_2$Te$_6$ (Y = Ge, Si), CrX$_3$ (X = Br, I), Cr$_5$Te$_8$ \cite{richter2018,casto2015,liu2019C5T8,pandey2024}. $T_C$ is determined from the minima of the temperature derivative of the magnetization curves $dM$ / $dT$ under FC conditions [inset of Fig. 1(c) and upper inset of 1(d)], yielding a value of about 78.4 K in a field of 200 Oe. The lower inset of Fig. 1(d) and Fig. S2 in the Supplementary Material shows the isothermal magnetization curves for both the $H$ $||$ $a$ and $H$ $||$ $bc$ orientations, measured up to a magnetic field of 7 T. The value of the saturation magnetization ($M_s$) estimated from the high field is around 1.1 $\mu_B$/Fe. From field-dependent magnetization measurements, the saturation field for $H$ $||$ $a$ is lower than that for $H$ $||$ $bc$, which indicates that the $a$-axis is the easy axis of magnetization. Furthermore, the observation of a narrow hysteresis loop suggests that NbFeTe$_2$ exhibits soft FM behavior.

AC magnetic susceptibility was employed to further elucidate the magnetic ground state. In-phase susceptibility in zero-field cooling, $\chi'(T)$, measured under an AC field of 2.5 Oe at frequencies between 10 Hz and 1 kHz, exhibits a sharp peak at 76 K, followed by a steep decline toward zero at lower temperatures [Fig. 1(e)]. The peak position is frequency independent, as highlighted in the upper inset of Fig. 1(e). The corresponding out-of-phase component, $\chi''(T)$, shown in the lower inset of Fig. 1(e), exhibits a pronounced peak near 73 K and falls abruptly to zero around 79 K. The maxima in $\chi''$($T$) also display no discernible frequency dependence, consistent with the behavior of a typical FM system. $\chi''$($T$) reflects the energy dissipation associated with the magnetization dynamics under an oscillating magnetic field, and it is proportional to the area of the dynamic hysteresis loop. Thus, $T_C$ is assigned to the temperature where $\chi''(T)$ vanishes above its maximum, i.e., $\sim$ 79 K, which corresponds well to the FM-PM transition. The frequency independence of both $\chi'(T)$ and $\chi''(T)$ maxima confirms the FM state. At lower temperatures, both components develop a shoulder-like anomaly around 50 K, although the exact positions and precise frequency dependence of the shoulders are difficult to determine. Unlike recent reports on NbFeTe$_2$ and Na-intercalated Cr$_2$Ge$_2$Te$_6$, where a frequency-dependent shift in the peak position of $\chi'(T)$ below $T_C$ is indicative of a spin-glass phase \cite{gao2025,khan2024}. Meanwhile, the frequency-independent nature of the peak positions of $\chi'(T)$ and $\chi''(T)$ in the present case, followed by the shoulder-like anomaly, and the abrupt downturn in $M_{ZFC}(T)$ at low field, suggests a possible low-temperature spin-glass state along with the FM ordering \cite{Dnam1999}. 

The magnetocrystalline anisotropy can be quantitatively estimated using the saturation field $H_s$, which contributes to the total micromagnetic energy density through the term $E_A =K_{u} \sin^2 (\theta-\phi)$, where $K_u$ is the uniaxial magnetocrystalline anisotropy constant, $\theta$ denotes the direction of the easy axis, and $\phi$ represents the actual direction of magnetization. The anisotropy energy reaches its maximum when $\theta - \phi = 90^{\circ}$, corresponding to the field applied within the $bc$-plane. The anisotropy constant $K_u$ can be determined using the relation $2K_u/M_s = \mu_0 H_s$, where $M_s$ is the saturation magnetization and $\mu_0$ is the vacuum permeability. The derived value of $K_u$ as a function of temperature, along with $H_s$($T$) and $M_s$($T$), is presented in the main panel and insets of Fig. 1(f). At 5 K, the value of $K_u$ is estimated to be 85.72 kJ m$^{-3}$, and it exhibits a monotonic decrease with increasing temperature, approaching negligible values in the vicinity of $T_C$. This reduction is likely driven by the thermal activation of local spin cluster fluctuations, which progressively suppress magnetic anisotropy at elevated temperatures. The presence of uniaxial anisotropy probably accounts for the pronounced downturn in the magnetization when the field is applied along the in-plane direction. The temperature-dependent decrease in magnetocrystalline anisotropy results in higher magnetization at elevated temperatures compared to lower temperatures in the low-field regime for $H$ $||$ $bc$ (Fig. S2 in the Supplemental Material), leading to a positive magnetic entropy change ($\Delta S_M$) as shown in Figs. S3(b) and (d) in the Supplemental Material. The positive  $\Delta S_M$ at low fields is likely associated with a canted FM spin configuration.

 \begin{figure*}
    \centering
     \includegraphics[width=1\linewidth]{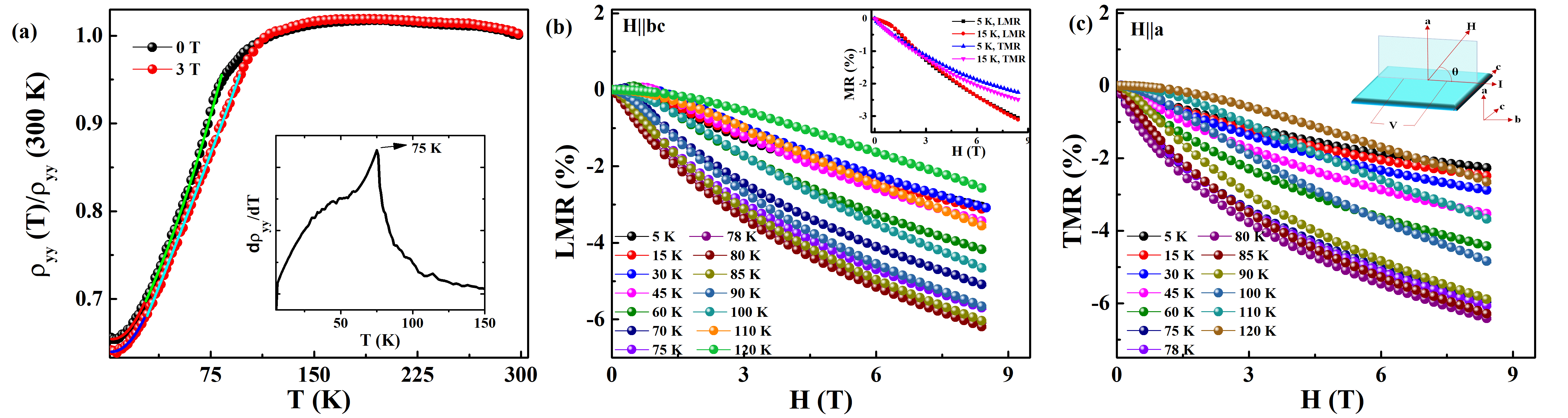}
    \caption{(a) Temperature dependence of the normalized longitudinal resistivity ($\rho_{yy}$) curves at zero-field and 3 T. Here, the current is applied along the in-plane direction. Theoretical fits to the temperature dependence of resistivity data at different temperature ranges are indicated with red, green, blue, and cyan colors. The inset shows the temperature derivative of resistivity. The magnetic field dependence of (b) LMR and (c) TMR at several representative temperatures up to 120 K. The inset of (b) shows the comparison between LMR (\%) and TMR (\%) at 5 K and 15 K. The measurement configuration for magnetoresistance is shown schematically in the inset of (c).}
    \label{fig:enter-label}
 \end{figure*}

 Fig. 2(a) illustrates the temperature dependence of the normalized longitudinal resistivity ($\rho_{yy}$) at 0 T and 3 T in the temperature range 2-300 K. The overall increasing nature of the resistivity with temperature confirms the metallic state of the system. A distinct change in slope near $T_C$ signals the onset of the FM transition, as seen more clearly in the $d\rho_{yy}/dT$ vs. $T$ curve, which exhibits a kink at 75 K. This value is in agreement with the $T_C$ determined from the magnetization measurements and reflects the suppression of spin disorder scattering in the FM state. The resistivity of metallic systems generally arises from multiple scattering processes \cite{Raquet2002,Jena2020}. To gain a deeper understanding of the underlying mechanisms, $\rho_{yy}(T)$ was analyzed in two distinct temperature regions. As shown in Fig. 2(a), the functional form of $\rho_{yy}(T)$ is highly sensitive to the temperature regime. In both regions, the low-temperature regime ($T <$ 23 K) and the intermediate region (23 to 80 K), the resistivity is best described by a combination of linear and quadratic terms in temperature. The linear term arises from the electron–phonon coupling, while the $T^2$ contribution is associated with the electron–magnon or electron–electron scattering. The pronounced field dependence of $\rho_{yy}(T)$ at 3 T indicates that electron–magnon scattering dominates over electron–electron scattering in both temperature ranges. However, the temperature coefficients obtained from the fits differ between the two regions, implying a change in the relative contributions of the scattering channels with temperature. The field dependence of $\rho_{yy}$ was examined by MR measurements, defined as $[\rho_{yy}(H)-\rho_{yy}(0)]/\rho_{yy}(0)$. To remove spurious Hall contributions caused by a slight misalignment between the electrodes, the data were symmetrized using $\rho_{yy}(H) = [\rho_{yy}(+H)+\rho_{yy}(-H)]/2$. Representative LMR curves, with $H\parallel I$, are shown in Fig. 2(b), while transverse MR (TMR), with $H\perp I$, is presented in Fig. 2(c). The measurement geometry is illustrated in the inset of Fig. 2(c), where the current was applied along an in-plane crystallographic axis, and the magnetic field was rotated from $\theta=0^\circ$ (LMR) to $\theta=90^\circ$ (TMR). The MR remains predominantly negative throughout the temperature range of 5 to 120 K, reaching a maximum near 80 K with values of 6.4\% for $H$ $ ||$ $a$ and 6.2\% for $H$ $||$ $bc$ at 8.5 T, and decreases with a further increase in temperature. In magnetic materials, negative MR generally originates from the suppression of electron–magnon scattering under an applied magnetic field. In contrast, in nonmagnetic systems (as well as magnetic materials hosting Weyl points), negative LMR is often attributed to the chiral anomaly. The chiral anomaly refers to the breakdown of chiral charge conservation at two Weyl nodes of opposite chirality in the presence of parallel electric (i.e., $I$) and magnetic fields, leading to charge pumping between the nodes \cite{SRoy2024}. However, in magnetic materials, disentangling the chiral-anomaly part from the FM contribution in the transport responses is challenging. Notably, as shown in the inset of Fig. 2(b), the absolute value of LMR (\%) remains higher than the TMR (\%) over the entire high-field range at low temperatures (up to $\sim$ 30 K), suggesting that, beyond a purely FM origin, additional mechanisms contribute to the longitudinal transport response in the low-temperature regime.
 
\begin{figure*}
    \centering
    \includegraphics[width=1\linewidth]{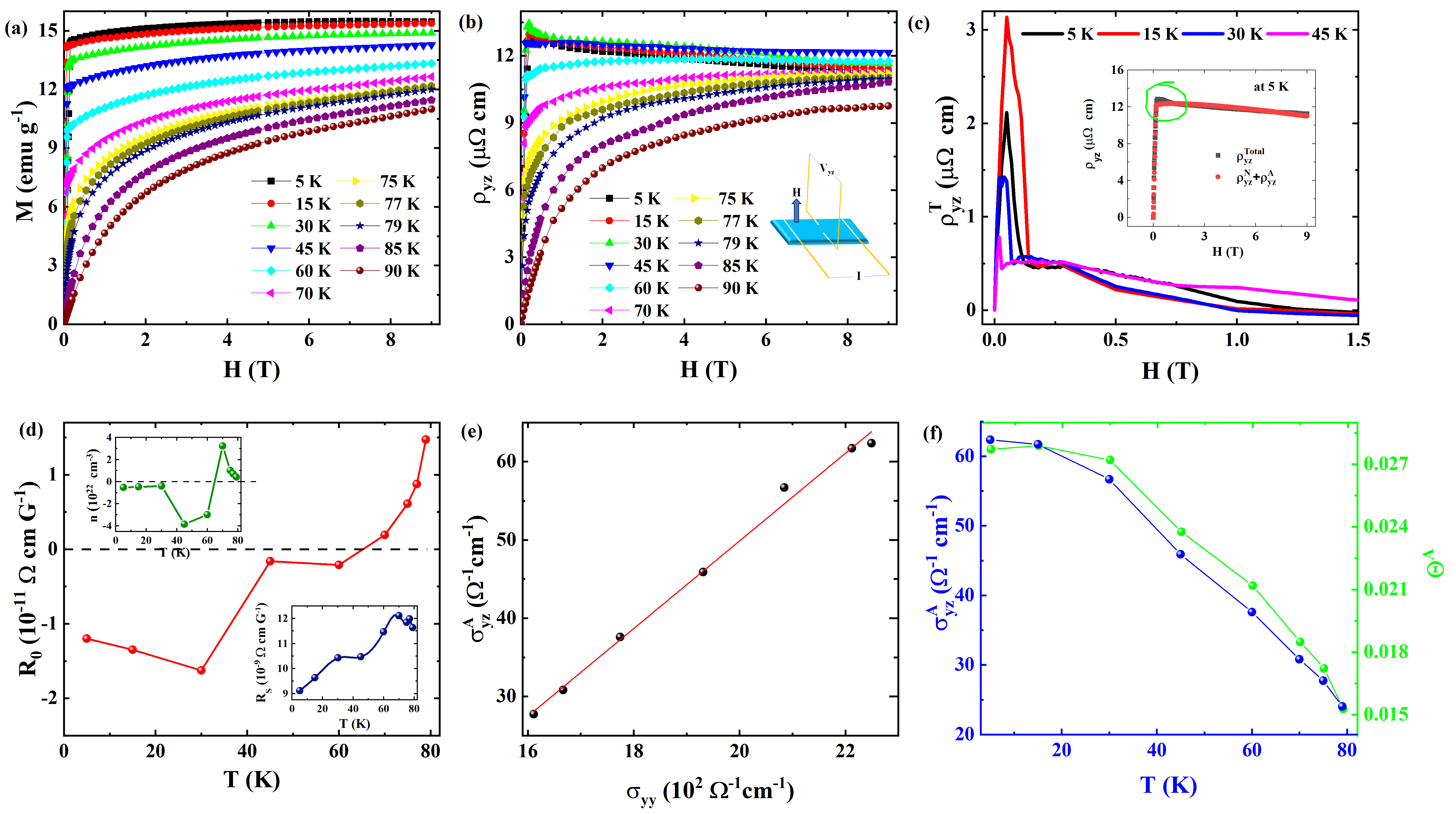}
    \caption{(a) Magnetic field dependence of magnetization ($M$) at different temperatures with field perpendicular to the $bc$-plane of the crystal. (b) Magnetic field dependence of Hall resistivity ($\rho_{yz}$) measured in the temperature range from 5 K to 90 K, with current in $bc$-plane and magnetic field along the $a$-axis of the crystal. (c) Field dependence of topological Hall resistance at different temperatures. The inset shows the total Hall resistivity $\rho_{yz}^{total}(H)$ and the sum of the ordinary Hall resistivity $\rho_{yz}^0(H)$ and anomalous Hall resistivity $\rho_{yz}^A(H)$ at 5 K. (d) Temperature dependence of the ordinary Hall coefficient ($R_0$). The upper inset shows the obtained carrier density as a function of temperature. The lower inset shows the temperature variation of the anomalous Hall coefficient $R_s$. (e) Scaling behavior of anomalous Hall conductivity $\sigma_{yz}^A$ vs longitudinal conductivity $\sigma_{yy}$ following a linear relation. (f) Temperature dependence of $\sigma_{yz}^A$ and anomalous Hall angle $\Theta^A$, both decrease with increasing temperature.}
    \label{fig:enter-label}
\end{figure*}
 
\begin{figure*} 
    \centering
    \includegraphics[width=1\linewidth]{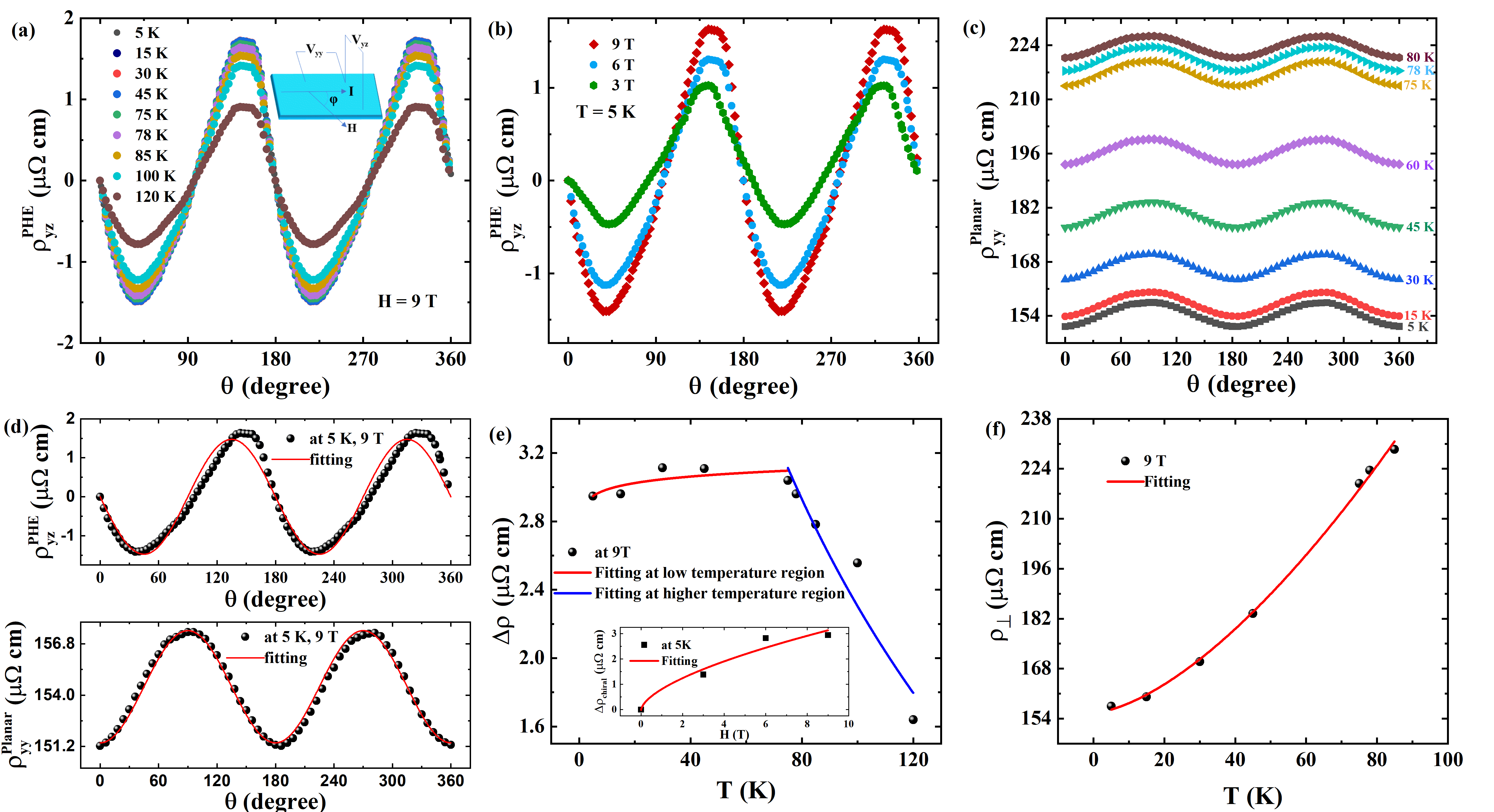}
    \caption{(a) The planar Hall resistivity ($\rho_{yz}^{\mathrm{PHE}}$) as a function of angle at 9 T for different temperatures. The inset schematic shows the experimental configuration or measurement set-up. (b) The $\rho_{yz}^{\mathrm{PHE}}$ for NbFeTe$_2$ as a function of angle at 5K for several magnetic fields. (c) The planar resistivity ($\rho_{yy}^{\mathrm{planar}}$) as a function of angle for different temperatures at 9T. (d) Global fittings of representative data for $\rho_{yz}^{\mathrm{PHE}}$ and $\rho_{yy}^{\mathrm{planar}}$ at 5 K and 9 T. (e) The extracted resistivity anomaly $\Delta\rho$ at 9 T as a function of temperature. Inset: The $\Delta\rho$ at 5 K as a function of magnetic field. (f) The extracted transverse resistivity $\rho_{\perp}$ as a function of temperature at 9T.}
    \label{fig:enter-label}
\end{figure*}

In FM systems, the evolution of transverse resistivity ($\rho_{yz}$) under an applied magnetic field is attributed to the anomalous Hall effect (AHE), which adds an additional contribution to the conventional Hall response. The total Hall resistivity can be expressed as \cite{Pugh1930,Pugh1932}: 
 $\rho_{yz} = \rho_{yz}^0 + \rho_{yz}^A = R_0 H + 4 \pi R_s M$, 
 where $\rho_{yz}^0$ and $\rho_{yz}^A$ denote the ordinary and anomalous Hall resistivity, respectively. Here, $R_0$ and $R_s$ are the ordinary and anomalous Hall coefficients. In the high-field regime, where the magnetization $M$ reaches saturation, the values of $R_0$ and $\rho_{yz}^A$ can be obtained from a linear fit to the $\rho_{yz}(H)$ data. The slope of this linear fit yields $R_0$, while the intercept on the $\rho_{yz}$ axis corresponds to $\rho_{yz}^A$. The saturation magnetization $M_S$ used in the analysis was extracted from the $M(H)$ curves at an applied field of 9 T, as shown in Fig. 3(a). $\rho_{yz}^0$ originates from the classical Lorentz force acting on the charge carriers and $\rho_{yz}^A$ is a manifestation of spin-dependent scattering and intrinsic Berry curvature effects in the FM phase. As illustrated in Fig. 3(b), field-dependent measurements of $\rho_{yz}$ conducted at fixed temperatures reveal a rapid increase in resistivity at low fields, which further asymptotically approaches a saturation value, characteristic of a pronounced anomalous Hall contribution below $T_C$. Interestingly, in the low field region, a small kink is observed in $\rho_{yz}$ vs. $H$ plot up to 45 K. This feature indicates the presence of an additional contribution beyond the ordinary and anomalous Hall effects, which we attribute to the topological Hall resistivity ($\rho_{yz}^T$). In magnetic systems hosting nontrivial spin textures, conduction electrons acquire a real-space Berry phase upon traversing these structures. This emergent Berry phase acts as an effective magnetic field, giving rise to the THE, a widely recognized transport signature of topological spin structures. Now, the total Hall resistivity takes the form 
$\rho_{yz} = \rho_{yz}^0 + \rho_{yz}^A + \rho_{yz}^T$. $\rho_{yz}^T$ was extracted by subtracting the fitted $(\rho_{yz}^0 + \rho_{yz}^A)$ from the total $\rho_{yz}$. As shown in Fig. 3(c), a finite THE is detected up to 45 K, appearing in the vicinity of the magnetic saturation. The maximum $\rho_{yz}^T$ reaches 3.1 $\mu\Omega\text{cm}$ near 0.05 T at 5 K and systematically decreases with increasing temperature. In this regime, the maximum value of $\rho_{yz}^T$ is comparable in magnitude to $\rho_{yz}^A$. Previous reports on Fe$_3$GeTe$_2$ and Fe$_3$GaTe$_2$ with out-of-plane magnetic anisotropy also exhibit THE, but the effect appears only when the magnetic field is applied parallel to the crystal plane, i.e., along the hard direction \cite{Algaidi2025,Roy2025}. A recent study on Cr$_{1.61}$Te$_2$ reported in-plane magnetic anisotropy and large THE, comparable to AHE, when the field is applied along the out-of-plane direction \cite{yhuang2025}. In this compound, the in-plane anisotropy favors the formation of microscopic noncoplanar spin configurations rather than mesoscopic skyrmionic textures, where the concentration of self-intercalated Cr-atoms significantly influences the magnetic spin structure. Furthermore, monoclinic Cr$_{1.53}$Te$_2$, with weak out-of-plane anisotropy, has been reported to host a sizable THE when the field is applied along the easy axis, showing distinct regimes associated with nontrivial spin textures at low temperature and skyrmions near room temperature \cite{czhang2023}. 

In NbFeTe$_2$, out-of-plane magnetic anisotropy is expected to support the stabilization of topological spin structures. However, the maximum value of $\rho_{yz}^T$, which is comparable to the value of $\rho_{yz}^A$, favors the formation of a noncoplanar spin structure. A detailed microscopic study and small-angle neutron scattering measurement will be crucial to elucidate the magnetic ground state and the underlying spin structure of NbFeTe$_2$. The M\"{o}sbauer spectroscopy study suggested a slight inhomogeneity in the magnetically ordered state in NbFeTe$_2$, which could arise from the variable magnetic interactions of Fe–Fe dumbbells with interstitial Fe atoms or with Nb atoms \cite{stepanova2024}. Density functional theory (DFT) calculations further indicate that the intralayer magnetic coupling is FM between Fe ions, whereas it is AFM between Fe and Nb ions \cite{wu2024}. However, the ground state of the interlayer coupling is FM \cite{wu2024}. Thus, competition between magnetic exchange interactions and uniaxial magnetic anisotropy promotes stabilization of intricate spin textures in NbFeTe$_2$. To facilitate a comparison of the relevant variables of different vdW ferromagnets exhibiting the THE reported in the literature alongside our present results, we compile the key parameters, including the temperature range over which THE was observed, magnetic anisotropy, maximum THE value, and the magnetic-field orientation associated with THE in Table I. The effective concentration of the charge carrier ($n$) can be estimated from the magnitude of $R_0$, assuming a single-band model. As illustrated in the main panel and the upper inset of Fig. 3(d), $R_0$ and $n$ undergo a sign reversal near 65 K, signifying a transition from hole-dominated to electron-dominated transport as the temperature decreases below this point. A similar temperature-dependent sign change in $R_0$ has also been observed in Fe$_5$GeTe$_2$ and Fe$_4$GeTe$_2$ \cite{PRM2019,npj2024}. The temperature dependence of $R_s$ is presented in the lower inset of Fig. 3(d), exhibits a non-monotonic trend, and reaches a maximum around 65 K before declining with further cooling. This behavior suggests a complex interplay between magnetic order and scattering mechanisms influencing the anomalous Hall effect across the temperature range. Furthermore, as shown in Fig. 3(e), the scaling between the anomalous Hall conductivity $\sigma_{yz}^A$ ($= \rho_{yz}^A/[(\rho_{yy})^2 + (\rho_{yz}^A)^2]$) and the longitudinal conductivity $\sigma_{yy}$ ($= \rho_{yy}/[(\rho_{yy})^2 + (\rho_{yz}^A)^2]$) below $T_C$ follows a linear equation which indicates that both intrinsic and extrinsic mechanisms contribute to AHE in this material. The anomalous Hall angle ( $\Theta^A$ $= \sigma_{yz}^A/\sigma_{yy}$), which quantifies the transverse current generated relative to the longitudinal current and $\sigma_{yz}^A$, both exhibit a monotonic decrease with increasing temperature, as shown in Fig. 3(f).

The planar Hall resistivity ($\rho_{yz}^{\mathrm{PHE}}$) of NbFeTe$_2$ as a function of angle at several temperatures under a 9 T magnetic field is shown in Fig. 4(a), with the corresponding measurement geometry illustrated in the inset. In the PHE configuration, the magnetic field, current, and Hall voltage lie in the same plane. The transverse and longitudinal voltages were measured by rotating the field ($H$) in the $bc$-plane, which corresponds to the planar Hall and the planar resistivity ($\rho_{yy}^{\mathrm{planar}}$), respectively. To eliminate spurious contributions from a conventional Hall voltage due to slight misalignment, the $\rho_{yz}^{\mathrm{PHE}}$ data were symmetrized with respect to positive and negative field directions. The angle dependence of $\rho_{yz}^{\mathrm{PHE}}$ and $\rho_{yy}^{\mathrm{planar}}$ can be described by \cite{Nandy2017,Burkov2017},
\begin{equation}
\rho_{yz}^{\mathrm{PHE}} = \Delta\rho\sin{\theta}\cos{\theta},
\label{eqn1}
\end{equation}
\begin{equation}
\rho_{yy}^{\mathrm{planar}} = \rho_{\perp} + \Delta\rho\cos^2{\theta}
\label{eqn2}
\end{equation}
\begin{equation}
\Delta\rho = \rho_{\|} - \rho_{\perp}, 
\label{eqn1}
\end{equation}
where $\theta$ is the angle between the current and the magnetic field, $\Delta\rho$ is the resistivity anomaly, $\rho_{\perp}$, and  $\rho_{\|}$ are the resistivity for transverse and longitudinal configurations, respectively. 
\begin{table*}
\centering
\caption{Overview of some key parameters: Curie temperature T$_C$, Temperature range exhibiting THE, Magnetic Anisotropy, Absolute Maximum value of THE ($|\mathrm{THE}_{\max}|$), Magnetic field orientation showing THE for different vdW ferromagnets.}
\vspace{0.2cm}
\setlength{\tabcolsep}{2pt}
\begin{tabular}{|c|c|c|c|c|c|c|}
\hline
 Material & T$_{C}$ (K) & \parbox[c]{3cm}{Temperature range\\exhibiting THE (K)} & Magnetic Anisotropy & \parbox[c]{3cm}{$|\mathrm{THE}_{\max}|$ ($\mu\Omega\text{cm}$)} & \parbox[c]{4cm}{ field orientation\\showing THE} & Reference\\
\hline
\rule{0pt}{20pt}Cr$_{1.61}$Te$_2$ & 326 & \parbox[c]{3cm}{75-260\\2-75} & in-plane & \parbox[c]{3cm}{0.93\\0.056}\quad & out-of-plane & \cite{yhuang2025}\\
\rule{0pt}{20pt}Cr$_{1.74}$Te$_2$ & 343 & 10-320 & out-of-plane & 0.88 & out-of-plane & \cite{Jliu2022}\\
\rule{0pt}{20pt}Cr$_5$Te$_8$ & 226 & 2-160 & out-of-plane & 0.16 & in-plane & \cite{ywang2019}\\ 
\rule{0pt}{20pt}Fe$_{5-x}$GeTe$_2$ & 275 & 5-260 & \parbox[c]{3cm}{in-plane (100-260 K)\\out-of-plane (T\textless100 K)}  & 0.74 & out-of-plane & \cite{YGao2020}\\
\rule{0pt}{20pt}Fe$_{3}$GaTe$_2$ & 359 & 150-350 & out-of-plane & 1.48 & in-plane & \cite{Algaidi2025}\\
\rule{0pt}{20pt}Fe$_{3}$GeTe$_2$ & 150 & 5-150 & out-of-plane & 2.04 & in-plane & \cite{You2019}\\ 
\rule{0pt}{20pt}NbFeTe$_2$ & 80 & 5-45 & out-of-plane & 3.1 & out-of-plane & this work\\ \hline
\end{tabular}
  \label{tab:my_label}
\end{table*}
The angular dependence of PHE at different temperatures (Fig. 4(a)) reveals maxima at 135$^{\circ}$ and 315$^{\circ}$ and minima at 45$^{\circ}$ and 225$^{\circ}$, in agreement with the expected PHE response of topological materials \cite{Nandy2017}. Figure 4(a) shows that the magnitude of $\rho_{yz}^{\mathrm{PHE}}$ increases with temperature up to 30 K before decreasing, although a finite PHE signal persists up to 120 K. $\rho_{yz}^{\mathrm{PHE}}$ as a function of angle at 5 K for several magnetic fields is shown in Fig. 4(b). The angle-dependent $\rho_{yy}^{\mathrm{planar}}$ for different temperatures in the 9 T field is shown in Fig. 4(c).  Using the above theoretical expressions, global fits of $\rho_{yz}^{\mathrm{PHE}}$ and $\rho_{yy}^{\mathrm{planar}}$ were performed, as shown in the upper and lower panels of Fig. 4(d). The extracted $\Delta\rho$ and $\rho_{\perp}$ are plotted as a function of temperature and magnetic field in Fig. 4(e) (main panel and inset) and 4(f). At 9 T, $\Delta\rho(T)$ shows a nonmonotonic behavior that initially increases with increasing temperature to $T_C$, followed by a rapid decrease with further increase of temperature. Meanwhile, $\rho_{\perp}$ shows an almost quadratic increase with temperature. At 5 K, $\Delta\rho$ rises systematically with increasing magnetic field and displays a power-law dependence with an exponent of 0.61. In general, PHE can arise from multiple microscopic mechanisms: (a) spin-dependent scattering in ferromagnets, where the magnetization orientation relative to the current regulates the spin scattering rates \cite{Parkin2020}, (b) anisotropic ordinary magnetoresistance (OMR) originating from anisotropy in Fermi pockets contributes to PHE \cite{Parkin2020}, (c) the chiral anomaly effect arising from charge pumping between Weyl nodes of opposite chirality \cite{xma2023}. In the present case, the similarity between the temperature dependence of $\Delta\rho(T)$ and the low-field $M(T)$ curve for $H$ $||$ $bc$ suggests that ferromagnetism contributes to the observed PHE. In contrast, for conventional itinerant FM systems, $\rho_{\perp}$ is smaller than $\rho_{\|}$, whereas $\rho_{yy}^{\mathrm{planar}}$ in NbFeTe$_2$ shows a clear deviation supporting the OMR-related origin of PHE. Interestingly, although the in-plane saturation field for this system is $\sim$ 1.5 T, the measured $\Delta\rho$ does not exhibit any saturation-like behavior even up to 9 T. If ferromagnetism were the primary origin of the observed PHE, $\Delta\rho$ would typically be expected to saturate once the magnetization reaches its in-plane saturation value. In addition, for an isotropic ferromagnet, $\Delta\rho$ is proportional to the square of the in-plane component of magnetization ($M_{\|}^2$) \cite{Anaz2008}. In NbFeTe$_2$, $\Delta\rho$ does not follow any systematic relation with $M_{\|}^2$. Thus, the persistence of the PHE well above $T_C$, together with the negative longitudinal magnetoresistance observed up to 120 K and the absence of saturation in $\Delta\rho$ even at high magnetic fields, indicates that mechanisms beyond simple FM alignment are responsible for the observed PHE. These features collectively suggest the presence of a possible nontrivial electronic band structure in this material.

Two recent independent studies have demonstrated that NbFeTe$_2$ synthesized at source temperatures in the range of 800–1000$^{\circ}$C stabilizes in a FM metallic ground state \cite{wu2024,stepanova2024}. The absence of frequency dependence in the peak positions of both the in-phase and out-of-phase components of ac susceptibility observed in our study is consistent with the reports above. In contrast, another recent work identified monoclinic NbFeTe$_2$, synthesized at a source temperature of 1000$^{\circ}$C under a modified temperature gradient, as an Anderson insulator exhibiting a spin-glass state below 0.7 T \cite{gao2025}. Such discrepancies can be attributed to differences in point defects, dislocations, and site disorder, which are strongly governed by synthesis conditions \cite{gao2025,May2016}. The associated lattice distortions play a decisive role in determining whether a metastable disordered magnetic state or an FM state is energetically favored. The spin-glass behavior reported in Ref. \cite{gao2025} suggests an enhanced disorder, likely arising from a chemical disorder or Nb vacancies that disrupt magnetic exchange pathways and thereby hinder the realization of the true intrinsic magneto-electronic ground state in monoclinic NbFeTe$_2$ at nominal stoichiometry \cite{Mcg2018}. The observed topological Hall effect (THE) in that system was attributed to locally broken inversion symmetry, which induces Dzyaloshinskii–Moriya (DM) interactions and stabilizes noncollinear spin textures around Nb-vacancy sites. In contrast, our investigation on monoclinic NbFeTe$_2$ crystals exhibiting a metallic FM behavior provides insight into the relationship between uniaxial magnetocrystalline anisotropy and topological phenomena. The observed topological Hall effect is suggested to arise from the competition between magnetic exchange interactions and out-of-plane anisotropy, which may facilitate the formation of nontrivial spin textures. Moreover, the presence of PHE along with negative LMR up to 120 K suggests the nontrivial electronic band structure in NbFeTe$_2$. Our findings establish this material system as a versatile platform for probing the intricate interplay between growth conditions, physical properties, magnetic anisotropy, and complex spin configurations.

\titlespacing*{\section}{0pt}{10pt}{10pt}
\section{CONCLUSIONS}
In summary, we have conducted a comprehensive investigation of the magnetic and transport properties of single crystal NbFeTe$_2$. The system exhibits a FM metallic behavior with $T_C$  \textasciitilde 80 K. Magnetization measurements reveal a pronounced out-of-plane magnetic anisotropy, with the anisotropy constant decreasing monotonically with increasing temperature. Magnetotransport measurements, performed under different field orientations, show negative LMR and TMR in the temperature range 5 to 120 K. When the field is applied along the out-of-plane direction perpendicular to the current, an additional kink in the Hall resistivity is observed near the magnetic saturation, which is attributed to the THE. The observation of THE in metallic NbFeTe$_2$ persisting up to 45 K points to the stabilization of intricate spin textures. Furthermore, a pronounced planar Hall effect, observed here for the first time, persists well above $T_C$ and, together with a negative LMR, indicates a possible nontrivial character of the underlying electronic band structure. These results establish NbFeTe$_2$ as a FM metallic system with complex spin dynamics and highlight its potential as a promising platform for spintronics and topological applications.

\section{ACKNOWLEDGMENTS}
The authors thank Dr. Nazir Khan for fruitful discussions and valuable comments. S.M. acknowledges SERB, Govt. of India, for funding through the National Post Doctoral Fellowship (PDF/2023/001273).

\bibliography{references}

\end{document}